\documentclass[aps,showpacs,twocolumn]{revtex4-2}
\usepackage[utf8]{inputenc}
\usepackage{amsfonts}
\usepackage{amssymb}
\usepackage{amsmath}
\usepackage{graphicx}
\usepackage{epsfig}
\usepackage{subfigure}
\usepackage{appendix}
\usepackage{hyperref}
\usepackage{color}

\hypersetup{hypertex=true, colorlinks=true, linkcolor=blue, urlcolor=blue, citecolor=blue}

\begin{document}

\title{Two-doublon Bloch oscillations in the mass-imbalanced extended
Fermi-Hubbard model}
\author{Kun-Liang Zhang}
\email{zhangkl@fosu.edu.cn}
\author{Xun-Da Jiang}
\author{Yong-Yao Li}
\affiliation{School of Physics and Optoelectronic Engineering, Foshan University, Foshan, 528225, China}
\affiliation{Guangdong-HongKong-Macao Joint Laboratory for Intelligent Micro-Nano Optoelectronic Technology, School of Physics and Optoelectronic Engineering, Foshan University, Foshan 528225, China}
\date{\today}

\begin{abstract}
Interactions between particles normally induce the decay of the particles Bloch oscillations (BOs) in a
periodic lattice. In the limit of strong on-site interactions, spin-$1/2$ fermions may form doublon bound states and undergo BOs in the presence of a tilted potential.
Here we investigate the impact of nearest-neighbor
interaction $V$ on the multi-doublon BOs in a mass-imbalanced extended
Fermi-Hubbard model. We derive an effective Hamiltonian for doublons, and show
that a slight change in $V$ can qualitatively alter their dynamic behaviors. Notably, at a resonance point, the doublons behave like free hard-core bosons. Under a tilted potential, the
system may exhibit different types of multi-doublon BOs at or deviation from the resonance point. Numerical results are presented to
demonstrate our conclusions in both one- and two-dimensional systems.
\end{abstract}

\maketitle

\section{Introduction}

\label{introduction}

Bloch oscillations (BOs) of electrons in crystal lattices were first predicted by Bloch in 1929 \cite{Bloch1929}. When a finite dc electric field is applied to the ideal crystal lattice, the eigenstates of electrons become localized, accompanied by the ladder-like energy levels, i.e., the Wannier-Stark (WS) ladder \cite{Wannier1960}, resulting in the coherent motion of wave packet, which is the underlying mechanism of the BOs. Sixty years after being predicted, the BOs were experimentally observed in semiconductor superlattices \cite{Feldmann1992, Waschke1993} and optical lattices \cite{Wilkinson1996, BenDahan1996, Anderson1998, Morsch2001}. In recent years, the BOs have been intensively explored in strongly correlated systems, including quantum spin chain \cite{kyriakidis1998bloch, cai2011quantum, shinkevich2012spectral, kosevich2013magnon, liu2019bloch, liu2021bloch, hansen2022magnetic, zhang2024magnetic}, interacting boson and fermion systems \cite{buchleitner2003interaction, freericks2008quenching, gustavsson2008control, eckstein2011damping, ribeiro2020many, scherg2021observing, claro2003interaction, dias2007frequency, dias2010role, kolovsky2004bloch, khomeriki2010interaction, kolovsky2010bose, longhi2012bloch, longhi2012correlated,  van2019bloch, longstaff2020bloch}, in both theoretical and experimental aspects.

In general,  interactions between particles induce the decoherence of  BOs \cite{buchleitner2003interaction, freericks2008quenching, gustavsson2008control, eckstein2011damping, ribeiro2020many, scherg2021observing}. Nevertheless, a strong interaction strength may bind two particles into a doublon \cite{strohmaier2010observation, sensarma2010lifetime, kajala2011expansion, longhi2012bloch, longhi2012correlated, hofmann2012doublon, longhi2013low, bello2017sublattice, rausch2017filling, garttner2019doublon, valmispild2020dynamically, cheng2024site} or a 
bound pair \cite{winkler2006repulsively, petrosyan2007quantum, jin2009coherent, kudo2009control, khomeriki2010interaction,  jin2011fast, kudo2011theoretical,  lin2014sudden} as high-energy excitations, in the boson and fermion systems, respectively. The BOs of bound particles in these systems exhibit characteristics different from the free-particle situations, for example, the  fractional BOs \cite{khomeriki2010interaction, corrielli2013fractional} or frequency doubling phenomena \cite{claro2003interaction, dias2007frequency, dias2010role, souza2010paired} have been observed. The Fermi-Hubbard model \cite{hubbard1963electron, arovas2022hubbard} describes a typical strongly correlated electronic system, with hopping amplitude $J$ and on-site interaction with strength $U$. A more general consideration leads to the extended Fermi-Hubbard model \cite{longhi2012bloch, hofmann2012doublon, esfahani2014nonlinear, van2015ultralong, wang2022experimental, qu2022spin, julia2024topological}, which includes the nearest-neighbor (NN) interaction with strength $V$. Obviously, both the on-site and NN interactions can affect the behavior of many-body BOs. 
The aforementioned research works have studied the BOs in the (extended) Fermi-Hubbard chains, but mainly focused on the one doublon or two fermions case. It is natural and interesting to take the doublon-doublon interaction into account in the dynamics of doublons BOs.

This paper aims to investigate the doublons BOs in the extended Hubbard model with imbalanced mass of spin-up and spin-down fermions \cite{longhi2013low, jin2015finite, grusha2016effective, sekania2017mass, sroda2019instability, heitmann2020density, zechmann2022tunable, darkwah2022probing, xie2023quench, kiely2024high}, under a tilted potential. We derive a doublons effective Hamiltonian in the strong on-site interaction situation to capture the physics of the doublons dynamics. Then we analyze the mechanism of BOs in the two-doublon subspace. We show that the NN interaction induces the bound-doublon eigenstates, which dominate the doublons dynamics. It is observed that (i) at a resonance point, the bound-doublon eigenstates disappear, and the doublons act like free hard-core bosons; (ii) when the NN interaction is strong enough, the energy levels of bound doublons are completely isolated from the scattering band of doublons. These properties enable the doublons BOs to occur after applying the tilted potential for the systems in the above two cases. To verify the analysis, we present the numerical simulations of time evolutions for the doublons dynamics, for the attractive on-site and NN interactions, i.e., $U<0$ and $V<0$, respectively, in the extended Fermi-Hubbard model. For the above two cases, the breathing modes of BOs exhibit the same pattern of doublon density, but with significant difference of mean-square displacements (MSD) of doublons in one dimension. The numerical results of doublons BOs for the system on a square lattice are also presented and discussed. Our results show that the multi-doublon BOs can survive in spite of the presence of the NN interaction of fermions.

The rest of this paper is organized as follows: In Sec. \ref{model}, we introduce the Hamiltonian of mass-imbalanced extended
Fermi-Hubbard model, and derive its doublons effective Hamiltonian. In Sec. \ref{BOs}, we analyze the mechanisms of the two types of doublons BOs, with the numerical results as the verifications. In Sec. \ref{BOs2D}, we discuss the doublons dynamics of the system on a square lattice. Finally, we conclude and discuss our findings in Sec. \ref{conclusion}.

\section{Model and doublons effective Hamiltonian}

\label{model}

\begin{figure}[t]
\centering
\includegraphics[width=0.5\textwidth]{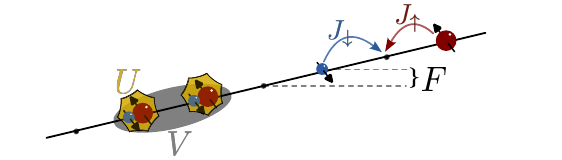}
\caption{Schematic illustration of the mass-imbalanced extended Fermi-Hubbard chain with 
the tilted potential $F$. Here the filled yellow polygons and gray ellipse represent
the on-site interaction $U$ and NN interaction $V$,
respectively. The hopping amplitudes for the heavy (red) and light (blue)
fermions are $J_{\uparrow}$ and $J_{\downarrow}$ ($J_{\downarrow}>J_{%
\uparrow}>0$), respectively. }
\label{fig_model}
\end{figure}

The Hamiltonian of the system we considered can be written as  
\begin{equation}
\mathcal{\hat{H}}=\hat{H}+\hat{H}_{\mathrm{dc}},  \label{H_total}
\end{equation}%
where 
\begin{eqnarray}
\hat{H} &=&-\sum_{\left\langle i,j\right\rangle }\sum_{\sigma =\uparrow
,\downarrow }J_{\sigma }\hat{c}_{i,\sigma }^{\dagger }\hat{c}_{j,\sigma }+%
\mathrm{H.c.}  \notag \\
&&+U\sum_{i}\hat{n}_{i,\uparrow }\hat{n}_{i,\downarrow }+\frac{1}{2}%
V\sum_{\left\langle i,j\right\rangle }\hat{n}_{i}\hat{n}_{j},
\label{H_hubbard}
\end{eqnarray}%
is the extended Fermi-Hubbard model with imbalance mass. Figure \ref{fig_model} presents a schematic illustration of the system in one dimension. For a one-dimensional ($1$D) system, the dc field term with strength $F$ takes the form of tilted potential 
\begin{equation}
\hat{H}_{\mathrm{dc}}=F\sum_{j}j\hat{n}_{j},
\label{H_dc}
\end{equation}%
which is readily feasible in the experiments of ultracold atoms \cite{guardado2020subdiffusion, scherg2021observing, kohlert2023exploring}.
Here, $\hat{c}_{i,\sigma }^{\dagger }$ ($\hat{c}_{i,\sigma }$) is the creation
(annihilation) operator of fermion with spin $\sigma $ on the site with index $i$%
; the summation $\sum_{\left\langle i,j\right\rangle }$ is over the
NN sites with indices $i$ and $j$; and $\hat{n}_{i,\sigma }=\hat{c}%
_{i,\sigma }^{\dagger }\hat{c}_{i,\sigma }$, $\hat{n}_{i}=\hat{n}%
_{i,\uparrow }+\hat{n}_{i,\downarrow }$ are the corresponding fermion number
operators. The strength of on-site interaction and NN
interaction are $U$ and $V$, respectively, which can be positive (repulsion)
or negative (attraction). The mass of the spin-up and spin-down fermions are
different, manifesting as the difference of the corresponding hopping amplitudes. This can 
be experimentally realized with Fermi gas, such as gas of fermionic atoms $^{171}\mathrm{Yb}$, $^{173}\mathrm{Yb}$, $^{87}\mathrm{Sr}$ or  $^{40}\mathrm{K}$, in the state-dependent optical lattices \cite{darkwah2022probing, riegger2018localized, heinz2020state, jotzu2015creating, bilokon2023many}. We assume $J_{\downarrow }>J_{\uparrow }>0$ in the
following discussions, and focus on the the attractive on-site and NN interactions case, i.e., $U<0$ and $V<0$, respectively. 
The range of on-site interaction strength $U/J_{\downarrow}\in [-20, 5]$ is experimentally accessible via magnetic-field tuning of the orbital Feshbach resonance, as, e.g., in the fermionic $^{171}\mathrm{Yb}$ atoms system \cite{darkwah2022probing}. 
It has been shown that the heavy particles acting as an effective disorder potential for the light ones in a extremely strong mass imbalance \cite{jin2015finite}. The system conserves the total fermion number $\hat{n}=\sum_{i}\hat{n}_{i}$, as well as the
total spin polarization $\hat{s}^{z}=\sum_{i}(\hat{n}_{i,\uparrow }-\hat{n%
}_{i,\downarrow })/2$. The imbalance of mass breaks the $\textrm{SU}(2)$ symmetry of the model, resulting in different propagation speeds of fermions with different spins in the lattice in general.

The standard Hubbard model in 1D was exactly solved by Lieb and Wu with the Bethe ansatz method \cite{lieb1968absence}.
The imbalance of mass $J_{\uparrow}\neq J_{\downarrow}$ and nonzero
NN interaction $V\neq 0$ resists the exact analytical treatments so far. 
Nevertheless, we pursue the perturbation and numerical solutions. In
this work, we focus on the dynamics of the system with multidoublon, i.e., each
site of the system is either doubly occupied or empty. This invariant
subspace, referred to as doublons subspace, is approximately achieved when
considering the strong on-site interaction limit $\left\vert U\right\vert
\gg J_{\sigma },\left\vert V\right\vert ,\left\vert F\right\vert $. In the
following, we derive the effective Hamiltonian of this case.

To explore the feasibility of the doublons BOs in the present model, we derive the doublons effective Hamiltonian. We firstly focus on the
Hamiltonian without the tilted potential. In the strong on-site interaction limit,
the states in the doublons subspace are well separated from others in the
spectrum. The doublons effective Hamiltonian for $\hat{H}$ can be obtained
perturbatively through the Schrieffer-Wolff transformation \cite{chao1977degenerate, chao1977kinetic, chao1978canonical, macdonald1988t, bravyi2011schrieffer}
\begin{eqnarray}
\hat{H}_{\mathrm{eff}} &=&e^{\mathrm{i}\hat{S}}\hat{H}e^{-\mathrm{i}\hat{S}}
\notag \\
&=&\hat{H}+\mathrm{i}\left[ \hat{S},\hat{H}\right] +\frac{\mathrm{i}^{2}}{2}%
\left[ \hat{S},\left[ \hat{S},\hat{H}\right] \right] +... \ ,
\end{eqnarray}%
where $\mathrm{i}=\sqrt{-1}$. The terms that change the number of doublons can be eliminated by properly
selecting the Hermitian operator $\hat{S}$. Up to second order, the doublons
effective Hamiltonian reads (see \ref{appendix_a} for detailed calculation) 
\begin{equation}
\hat{H}_{\mathrm{eff}}=\sum_{\left\langle i,j\right\rangle }\left[ J_{%
\mathrm{eff}}\left( \hat{d}_{i}^{\dagger }\hat{d}_{j}+\hat{d}_{j}^{\dagger }%
\hat{d}_{i}\right) +\frac{1}{2}V_{\mathrm{eff}}\hat{n}_{i}\hat{n}_{j}\right],
\label{H_eff}
\end{equation}%
where $d_{i}^{\dagger }=\hat{c}_{i,\uparrow }^{\dagger }\hat{c}%
_{i,\downarrow }^{\dagger }$ is the creation operator of doublon; the effective hopping amplitude and NN interaction of doublons are $J_{%
\mathrm{eff}}=2J_{\uparrow }J_{\downarrow }/U$ and $V_{\mathrm{eff}%
}=V-(J_{\uparrow }^{2}+J_{\downarrow }^{2})/U$, respectively. 
This means that the doublons move in the lattice through the second-order processes, and all three parameters of the Hubbard model play a crucial part in the doublon-doublon interaction.  
Clearly, the doublons
satisfies the statistics of hard-core bosons, that is 
\begin{eqnarray}
\left\{ \hat{d}_{i},\hat{d}_{i}^{\dagger }\right\} &=&1,\left\{ \hat{d}_{i},%
\hat{d}_{i}\right\} =0,  \notag \\
\left[ \hat{d}_{i},\hat{d}_{j}^{\dagger }\right] &=&\left[ \hat{d}_{i},\hat{d%
}_{j}\right] =0,(i\neq j).
\end{eqnarray}
That is to say, $\hat{H}_{\mathrm{eff}}$ in Eq. (\ref{H_eff}) suggests that
in the strong on-site interaction region, the nonequilibrium dynamics of
doublons in the mass-imbalanced extended Hubbard model Eq. (\ref{H_hubbard})
resembles that in a hard-core bosons system with effective hopping amplitude 
$J_{\mathrm{eff}}$ and NN interaction $V_{\mathrm{eff}}/2$. 
This is distinguished from the half-filled ground state in the case of $U>0$, the effective
Hamiltonian of which is the spin-$1/2$ anisotropic XXZ Heisenberg chain \cite{grusha2016effective},
and the introducing of NN interaction $V$ only gives a constant term 
in that. Notably, the above result is independent of spatial dimension.

\section{Doublons Bloch oscillations}

\label{BOs}

We have seen that the NN interaction $V$ plays a vital role in the
doublon-doublon interaction from the doublons effective Hamiltonian Eq. (\ref{H_eff}).
It is worth noting that when the energies of interactions and hoppings
resonate, the doublon-doublon interaction vanishes. In this section, we
investigate the doublons BOs under the tilted potential $F$ in two parameter
regions.

\subsection{Two-doublon Bloch oscillations}

In order to gain insight into the doublon dynamics under the tilted potential, we
first analyze the two-doublon situation in a chain with length $L$. Due to the translational invariance of the lattice, the
effective Hamiltonian Eq. (\ref{H_eff}) in two-doublon subspace can be written as 
\begin{eqnarray}
\hat{H}_{\mathrm{eff}}^{(2)}(K) &=&2J_{\mathrm{eff}}\cos \left( \frac{K}{2}%
\right) \sum_{r=1}^{L-1}\left[ \left\vert \phi _{r}(K)\right\rangle
\left\langle \phi _{r+1}(K)\right\vert +\mathrm{H.c.}\right]   \notag \\
&&+2V_{\mathrm{eff}}\left\vert \phi _{1}(K)\right\rangle
\left\langle \phi _{1}(K)\right\vert ,
\label{H_eff_K}
\end{eqnarray}%
under the two-doublon basis with momentum $K$ 
\begin{equation}
\left\vert \phi _{r}(K)\right\rangle =\frac{1}{\sqrt{L-1}}e^{\mathrm{i}%
Kr/2}\sum_{j}e^{\mathrm{i}Kj}\hat{d}_{j}^{\dagger }\hat{d}_{j+r}^{\dagger
}\left\vert \text{\textrm{Vac}}\right\rangle ,
\end{equation}%
where the vacuum state is defined as $\hat{d}_{i}\left\vert \text{\textrm{Vac%
}}\right\rangle =0$, and $r=1,2,3,...,L-1$ is the distance between two
doublons. Note that $\hat{H}_{\mathrm{eff}}^{(2)}(K)$ can be regarded as the
single-particle Hamiltonian of a tight-binding chain with hopping amplitude $%
2J_{\mathrm{eff}}\cos (K/2)$, and with impurity potential $2V_{\mathrm{eff}}$ 
at one end. Such impurity model supports bound state in certain parameter
region. In order to solve the eigenvalue equation $\hat{H}_{\mathrm{eff}%
}^{(2)}(K)\left\vert \psi (K)\right\rangle =E(K)\left\vert \psi
(K)\right\rangle $, we expand the eigenstate $\left\vert \psi
(K)\right\rangle $ as 
\begin{equation}
\left\vert \psi (K)\right\rangle =\sum_{r=1}^{L-1}C_{r}(K)\left\vert \phi
_{r}(K)\right\rangle .
\end{equation}%
Substituting this into the eigenvalue equation, we obtain 
\begin{eqnarray}
&&2J_{\mathrm{eff}}\cos \left( \frac{K}{2}\right) \left[
C_{r+1}(K)+C_{r-1}(K)\right]   \notag \\
&&+\left[ 2V_{\mathrm{eff}}\delta _{r,1}-E(K)\right] C_{r}(K)=0,
\label{eigen_eq}
\end{eqnarray}%
with boundary condition being $C_{0}(K)=0$. All the eigenstates and
eigenenergies of Hamiltonian Eq. (\ref{H_eff_K}) can be obtained by solving this
equation, numerically or analytically. 

\begin{figure}[t]
\centering
\includegraphics[width=0.5\textwidth]{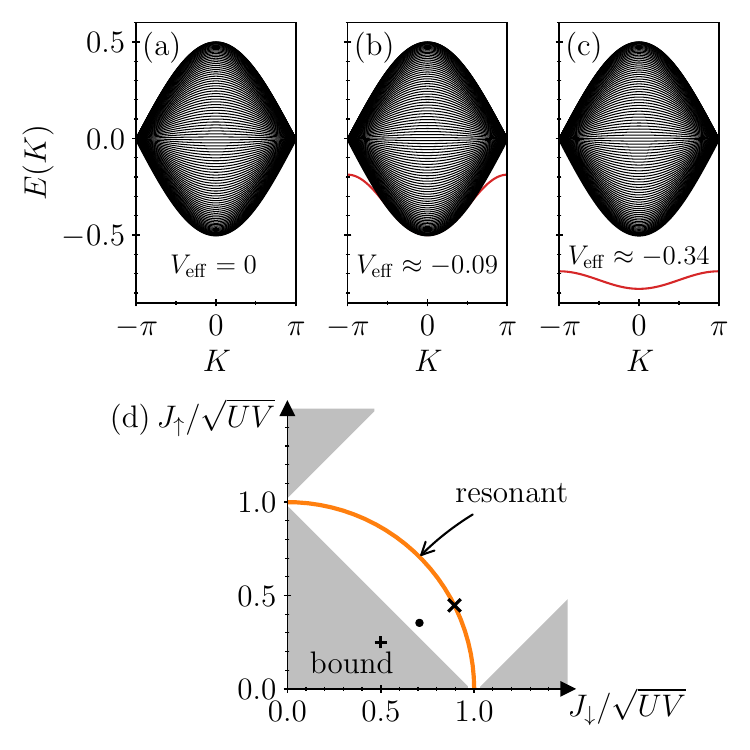}
\caption{Two-doublon spectrum as a function of momentum $K$ for various
NN interactions, which are obtained from the exact diagonalization of the two-doublon effective
Hamiltonian in Eq. (\ref{H_eff_K}). Parameters of the systems are (a) $V=-5/32$, (b) $V=-0.25$ and
(c) $V=-0.5$. Other parameters are $L=100$, $U=-8$, $J_{\downarrow}=1$ and $%
J_{\uparrow}=0.5$. The effective NN interactions $V_{\textrm{eff}}$ are marked in each panels, and the effective hopping amplitudes are all $J_{\textrm{eff}}=-0.125$. The doublons scattered bands and bound-doublon energy levels are indicated by the black and red lines, respectively. Here the energy is in units of $J_{\downarrow}$. (d) Parameter regions of hopping amplitudes where the systems support bound doublons BOs (gray area) and resonant doublons BOs (orange line). The markers `$\times$", ``$\cdot$" and ``$+$" indicate the regions  of parameters for (a), (b) and (c), respectively.  Here the unit is set as $\sqrt{UV}$, and strong on-site interaction and $U<0$, $V<0$ are assumed.}
\label{fig_spectrum}
\end{figure}

Next we focus on the bound-doublon
solution. We takes the ansatz for the coefficient of the bound state as $%
C_{r}(K)=\exp [-\beta (K)r]$ with $\beta (K)>0$ being the decay rate of wave function.
Substituting it into Eq. (\ref{eigen_eq}), straightforward calculation gives
the decay rate 
\begin{equation}
\beta (K)=\ln \left[ \frac{V_{\mathrm{eff}}}{J_{\mathrm{eff}}\cos \left(
K/2\right) }\right] ,
\end{equation}%
and the bound-doublon energy 
\begin{equation}
E_{\textrm{bou.}}(K)=2 V_{\mathrm{eff}}+\frac{2J_{\mathrm{eff}}^{2}\cos ^{2}\left(
K/2\right) }{V_{\mathrm{eff}}}.
\end{equation}%
The bound-state condition $\beta (K)>0$ requires $\left\vert V_{\mathrm{eff}%
}\right\vert >\left\vert J_{\mathrm{eff}}\cos \left( K/2\right) \right\vert $. In this case, the doublons bound states are induced by the NN interaction 
$V$. In Fig. \ref{fig_spectrum}, we present the two-doublon spectrum for various NN interaction strengths, which can be obtained by diagonalizing the two-doublon effective Hamiltonian Eq. (\ref{H_eff_K}) or solving Eq. (\ref{eigen_eq}). In Fig. \ref{fig_spectrum}(a), we can see that when $V_{\mathrm{eff}}=0$, the spectrum is all composed of scattered band of doublons. Increasing $\left| V \right|$ would induce the isolated energy levels, which are the energies of bound doublons [see Fig. \ref{fig_spectrum}(b)]. When $\left\vert V_{\mathrm{eff}}\right\vert >\left\vert J_{\mathrm{eff}}\cos \left( K/2\right) \right\vert $ for all $K$, that is, $\left\vert V_{\mathrm{eff}}\right\vert >\left\vert J_{\mathrm{eff}}\right\vert $,  the bound-doublon band is completely isolated from the scattered band [see Fig. \ref{fig_spectrum}(c)]. We illustrate these two parameter regions in Fig. \ref{fig_spectrum}(d) in units of $\sqrt{UV}$. In the following, we show that the systems in the two parameter regions support bound and resonant doublons BOs, respectively.

Here we first focus on the dynamics of the two bound doublons. For a
large NN interaction $V$, we have $\left\vert V_{\mathrm{eff}}\right\vert
\gg \left\vert J_{\mathrm{eff}}\cos \left( K/2\right) \right\vert $. In this case, two doublons in
the bound band are strongly bound. Then we are able to write
down the two-doublon effective Hamiltonian under the strongly bound basis $%
\left\vert \varphi _{j}\right\rangle =\hat{d}_{j}^{\dagger }\hat{d}%
_{j+1}^{\dagger }\left\vert \text{\textrm{Vac}}\right\rangle $. We divide
the effective Hamiltonian Eq. (\ref{H_eff}) into two parts, i.e., $\hat{H}_{\mathrm{eff%
}}^{(0)}=V_{\mathrm{eff}}\sum_{j}\hat{n}_{j}\hat{n}_{j+1}/2$ and $\hat{H}_{\mathrm{eff}%
}^{\prime }=J_{\mathrm{eff}}\sum_{j}( \hat{d}_{j}^{\dagger }\hat{d}_{j+1}+\hat{d}%
_{j+1}^{\dagger }\hat{d}_{j}) $, and treat $\hat{H}_{\mathrm{eff}%
}^{\prime }$ as a perturbation.  According to the degenerate perturbation
theory, the effective Hamiltonian is given by the following formula 
\begin{eqnarray}
\hat{H}_{\mathrm{eff}}^{(2,\mathrm{bou.})} &=&\hat{P}\hat{H}_{\mathrm{eff}%
}^{(0)}\hat{P}+\hat{P}\hat{H}_{\mathrm{eff}}^{\prime }\hat{Q}\frac{1}{E_{%
\mathrm{eff}}^{(0)}-\hat{H}_{\mathrm{eff}}^{(0)}}\hat{Q}\hat{H}_{\mathrm{eff}%
}^{\prime }\hat{P}\notag \\
&&+O\left( J_{\mathrm{eff}}^{3}/V_{\mathrm{eff}}^{2}\right) ,
\label{perturbation_H}
\end{eqnarray}
up to second order $J_{\mathrm{eff}}^{2}/V_{\mathrm{eff}}$. Here the
projectors are defined as $\hat{P}=\sum_{j}\left\vert \varphi
_{j}\right\rangle \left\langle \varphi _{j}\right\vert $ and $\hat{Q}=\hat{1}%
-\hat{P}$, and the energy of the unperturbed state is $E_{\mathrm{eff}%
}^{(0)}=2V_{\mathrm{eff}}$. Direct calculation shows that 
\begin{equation}
\hat{H}_{\mathrm{eff}}^{(2,\mathrm{bou.})}\approx \frac{J_{\mathrm{eff}}^{2}%
}{2V_{\mathrm{eff}}}\sum_{j}\left( \left\vert \varphi _{j}\right\rangle
\left\langle \varphi _{j+1}\right\vert +\left\vert \varphi
_{j+1}\right\rangle \left\langle \varphi _{j}\right\vert \right) ,
\label{H_eff_2bound}
\end{equation}
in which the constant potential that is proportional to $\sum_{j}\left\vert \varphi
_{j}\right\rangle \left\langle \varphi _{j}\right\vert$ is omitted. Now the bound-doublon physics is clear. From Eq. (\ref{H_eff_2bound}), we can see the  bound-doublon hops in the lattice through a  second-order process of the single-doublon hopping.

\begin{figure}[t]
\centering
\includegraphics[width=0.5\textwidth]{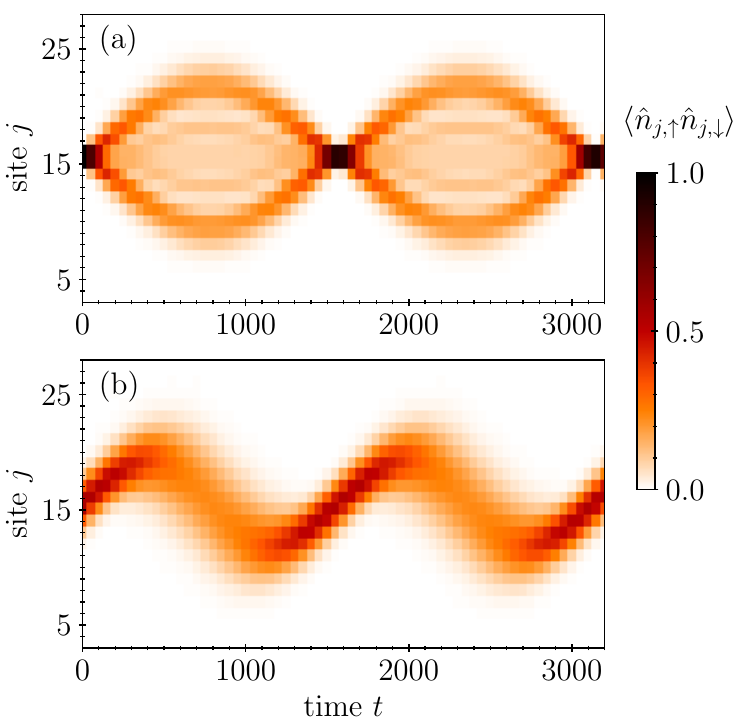}
\caption{Numerical results of the time evolution of doublon density defined in Eq. (\ref{dou_den}) for different initial states with (a) delta distribution in Eq. (\ref{ini_sta_a}) and (b) Gaussian distribution in Eq. (\ref{ini_sta_b}).
Parameters of the system are taken as $L=30$, $U=-6$, $V=-2$, $J_{\downarrow}=1$ and $%
J_{\uparrow}=0.5$. The effective NN interaction is $V_{\textrm{eff}}\approx -1.79$, and the effective hopping amplitude is $J_{\textrm{eff}}\approx -0.167$. The strength of tilted potential is taken as $F=0.001$. Parameters of the Gaussian wave packet are $\alpha=0.4$, $j_{0}=15$ and $K_{0}=\pi/2$. The time unit is $J_{\downarrow}^{-1}$, and we set $J_{\downarrow}=1$ throughout the numerical simulations in this work.}
\label{fig_boundBO}
\end{figure}

\begin{figure*}[t]
\centering
\includegraphics[width=1\textwidth]{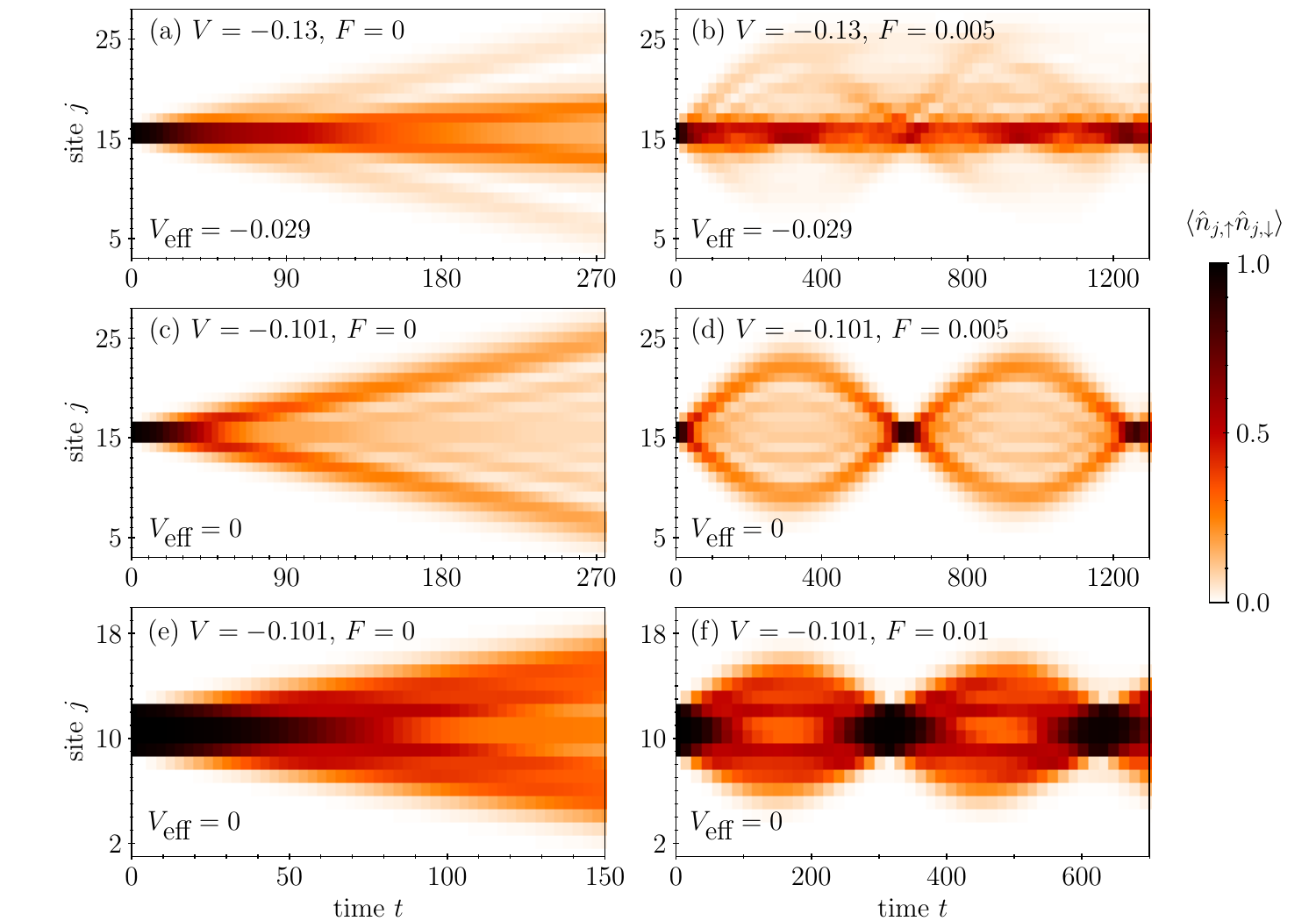}
\caption{Numerical results of the time evolution of doublon density for different NN interaction $V$, tilted potential $F$ and doublon number. The system   sizes for the two-doublon cases in (a)-(d) and four-doublon cases in (e) and (f) are taken as $L=30$ and $L=20$, respectively.  Other
system parameters are $U=-10$, $J_{\downarrow}=1$ and $J_{\uparrow}=0.1$. The effective NN interactions $V_{\textrm{eff}}$ are marked in each panels, and the effective hopping amplitudes are all $J_{\textrm{eff}}=-0.02$.} 
\label{fig_DoublonsBOs}
\end{figure*}

When the tilted potential $\hat{H}_{\mathrm{dc}}$ in Eq. (\ref{H_dc}) is applied, the quadrupled dc field $4F\sum_{j}j\left\vert \varphi
_{j}\right\rangle \left\langle \varphi _{j}\right\vert$ arises in the bound-doublon subspace, forming the WS ladder for bound doublons. Then the bound doublons may undergo BOs. The oscillating period would be $T=2\pi/(4F)$, which is a quarter of that of a free particle in lattice \cite{hartmann2004dynamics}. To verify the above analysis, we conduct numerical simulations of time evolutions. We consider the delta and Gaussian distributions of initial bound-doublon states, which are 
\begin{equation}
	\left\vert \Psi_{\mathrm{a}} \left( 0 \right) \right> =\hat{d}_{j_{0}}^{\dagger }\hat{d}%
_{j_{0}+1}^{\dagger }\left\vert \text{\textrm{Vac}}\right\rangle,
\label{ini_sta_a}
\end{equation}
and
\begin{equation}
	\left\vert \Psi_{\mathrm{b}} \left( 0 \right) \right> =\Omega^{-1}\sum_{j}e^{-\alpha^{2}(j-j_{0})^{2}+\mathrm{i}K_{0}j}\hat{d}_{j}^{\dagger }\hat{d}%
_{j+1}^{\dagger }\left\vert \text{\textrm{Vac}}\right\rangle,
\label{ini_sta_b}
\end{equation}
respectively. Here $\Omega$ is the normalization constant, $\alpha$ characterizes the width, $j_{0}$ is the center, and $K_{0}$ is the wave vector of the wave packet. The evolved state under the Hamiltonian in Eq. (\ref{H_total}) has the form $\left\vert \Psi \left( t \right) \right>=\exp(-\mathrm{i}\mathcal{\hat{H}}t)\left\vert \Psi \left( 0 \right) \right>$, which can be accurately computed by the Chebyshev expansion method \cite{tal1984accurate, fehske2009numerical, weisse2006kernel} for a relatively large system size, with the dimension of Hilbert space up to about $2\times 10^{7}$ in this work. The detail of the implementation of this method is presented in \ref{appendix_b}. The subsequent numerical simulations of time evolutions are all performed in this scheme. After obtaining the evolved state, we compute the doublons density for different lattice sites and time, which is defined as 
\begin{equation}
	\left< \hat{n}_{j,\uparrow}\hat{n}_{j,\downarrow} \right>=\left\langle \Psi \left( t \right)\right\vert \hat{n}_{j,\uparrow}\hat{n}_{j,\downarrow} \left\vert \Psi \left( t \right) \right>.
	\label{dou_den}
\end{equation}
Experimentally, the doublons density and its correlation can be directly measured in the Fermi-Hubbard lattice gas through the spin- and density-resolved quantum gas microscopy \cite{mitra2018quantum, koepsell2020robust, hartke2023direct}. 
The numerical results are presented in Fig. \ref{fig_boundBO}, which are consistent with our analysis. Figures. \ref{fig_boundBO}(a) and (b) show that two different initial excitations undergo the  breathing mode and oscillating mode, respectively. The oscillating periods are both $T=2\pi/(4F)\approx 1570.80 $.

\begin{figure}[t]
\centering
\includegraphics[width=0.5\textwidth]{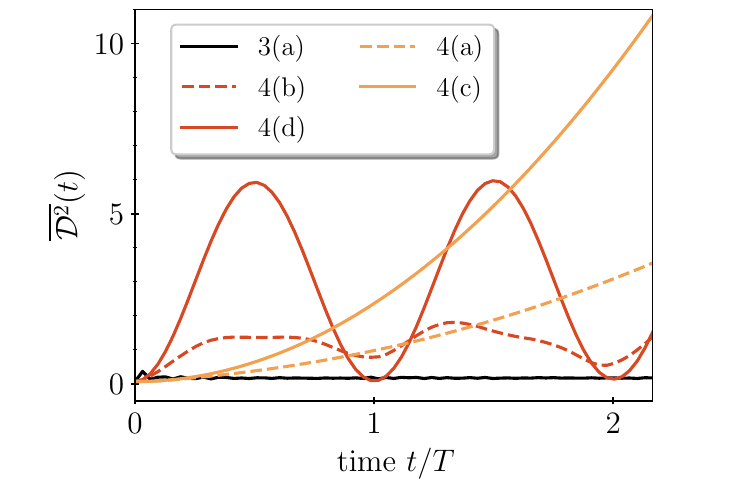}
\caption{Mean-square displacement between the doublons [see Eq. (\ref{avg_D})] as a function of time for the cases in Fig. \ref{fig_boundBO}(a) with parameters $U=-6$, $V=-2$, $J_{\downarrow}=1$, $J_{\uparrow}=0.5$, $F = 0.001$, $V_{\textrm{eff}}\approx -1.79$ and $J_{\textrm{eff}}\approx -0.167$. and Figs. \ref{fig_DoublonsBOs}(a)-(d) with parameters $U=-10$, $J_{\downarrow}=1$, $J_{\uparrow}=0.1$ and $J_{\textrm{eff}}= -0.02$. While $V=-0.13$, $F = 0$, $V_{\textrm{eff}}=-0.029$ for Fig. \ref{fig_DoublonsBOs}(a); $V=-0.13$, $F = 0.005$, $V_{\textrm{eff}}=-0.029$ for Fig. \ref{fig_DoublonsBOs}(b); $V=-0.101$, $F = 0$, $V_{\textrm{eff}}=0$ for Fig. \ref{fig_DoublonsBOs}(c); $V=-0.101$, $F = 0.005$, $V_{\textrm{eff}}=0$ for Fig. \ref{fig_DoublonsBOs}(d). The system sizes are all $L=30$. The time scales are normalized by the final time of the data in each case.}
\label{fig_MSD}
\end{figure}

\subsection{Free doublons dynamics in an interacting system}
Now we investigate the BOs at the resonance point of system parameters, which is $V_{\mathrm{eff}%
}=V-(J_{\uparrow }^{2}+J_{\downarrow }^{2})/U=0$. We are interested in the multi-doublon dynamics. Obviously, the released doublons initial state acts like free hard-core bosons according to the doublons effective Hamiltonian in Eq. (\ref{H_eff}), but with renormalized NN hopping amplitude $J_{\mathrm{eff}}$. Apart from the resonance nature, it is also worth noting that the doublons BOs remain, despite the presence of hard-core constraint. 
In Figs. \ref{fig_DoublonsBOs}(a) and (c), we present the numerical results of time evolution for the initial state in Eq. (\ref{ini_sta_a}) with different NN interaction but without the tilted potential. It is shown that the deviation from the resonance condition $V=(J_{\uparrow }^{2}+J_{\downarrow }^{2})/U$ obstructs the transport of doublons. This can be demonstrated by the following correlation function that characterize the MSD \cite{lev2017transport} between the doublons
 \begin{equation}
 	\overline{\mathcal{D}^{2}}(t) =\frac{1}{L}\sum_{i,j}(i-j)^{2}\left\langle \Psi \left( t \right)\right\vert \hat{n}_{i,\uparrow}\hat{n}_{i,\downarrow}  \hat{n}_{j,\uparrow}\hat{n}_{j,\downarrow}\left\vert \Psi \left( t \right) \right>.
 	\label{avg_D}
 \end{equation}
The definition of MSD involves the doublon density correlator, which can be measured through the quantum gas microscopy \cite{mitra2018quantum}.
The numerical results of the MSDs for these two cases are presented in Fig. \ref{fig_MSD}. For the resonant case, we can observe the characteristic of ballistic transport, i.e., $\overline{\mathcal{D}^{2}}(t) \sim t^{2}$, which means the doublons spread without scattering.

When applying the tilted potential in Eq. (\ref{H_dc}), the doublons effective Hamiltonian becomes 
\begin{equation}
\mathcal{\hat{H}}_{\mathrm{eff}}= J_{\mathrm{eff}}\sum_{j}\left( \hat{d}_{j}^{\dagger }\hat{d}_{j+1}+\hat{d}_{j+1}^{\dagger }%
\hat{d}_{j}\right)+2F\sum_{j}j\hat{d}_{j}^{\dagger }\hat{d}_{j},
\end{equation}%
where the doubling of the tilted potential comes form the double occupancy of fermions per site. This yields the WS ladder for free doublons. 
 The period of the BOs would be a half of that of free particle in lattice. The numerical result in Fig. \ref{fig_DoublonsBOs}(d) shows the free-doublon BOs at the resonance point. From Figs. \ref{fig_DoublonsBOs}(a), \ref{fig_DoublonsBOs}(b) and the corresponding MSDs in Fig. \ref{fig_MSD}, we can see that a minor deviation from the parameter $V=(J_{\uparrow }^{2}+J_{\downarrow }^{2})/U$ induces decoherence of the diffusion and BO, highlighting the effect of NN interaction. This can be grasped from the values of the effective parameters, which dominate the doublons dynamics. Although the deviation of $V$ is minor in comparison to $J_{\sigma}$ and $U$, the deviation of the effective NN interaction $\Delta V_{\textrm{eff}}=0.029$ is comparable with the effective hopping amplitudes $J_{\textrm{eff}}=-0.02$. 
 In Figs. \ref{fig_DoublonsBOs}(e) and (f), we present the results for the initial state with $4$ doublons ($8$ fermions)
 \begin{equation}
 	\left\vert \Psi \left( 0 \right) \right> =\prod_{i=9}^{12}\hat{d}_{i}^{\dagger }\left\vert \text{\textrm{Vac}}\right\rangle,
 \end{equation}
  corresponding to Figs. \ref{fig_DoublonsBOs}(c) and (d), but with a smaller system size $L=20$. The larger tilted potential in Fig. \ref{fig_DoublonsBOs}(f) is taken to constrain the amplitude of BO, avoiding the boundary effect. The diffusion behavior and BO are similar with the cases of $2$ doublons. The oscillating periods for the breathing modes are $T=2\pi/(2F)\approx 628.32 $ and $314.16$ for the cases in Figs. \ref{fig_DoublonsBOs}(d) and \ref{fig_DoublonsBOs}(f), respectively.
 
 We would like to stress that the breathing modes of BOs in Fig. \ref{fig_boundBO}(a) and Fig. \ref{fig_DoublonsBOs}(d) are different, although the profiles of their  doublons densities look almost the same. Firstly, as we have pointed out, their oscillating periods are a quarter and a half of that of free particle in lattice, respectively. Secondly, in Fig. \ref{fig_boundBO}(a), two doublons are always bounded tightly together by the NN interaction, evolving as a bound-doublon wave packet, while in Fig. \ref{fig_DoublonsBOs}(d), the distance between two doublons changes periodically since they can be considered as free particles. 
 The numerical results of the MSDs $\mathcal{D}(t)$ in Fig. \ref{fig_MSD} verify the above analysis. Nevertheless, one common point is that the spin-up and spin-down fermions are always well bounded as doublons,  regardless of the significant difference in their mass.

\begin{figure}[tbh]
\centering
\includegraphics[width=0.5\textwidth]{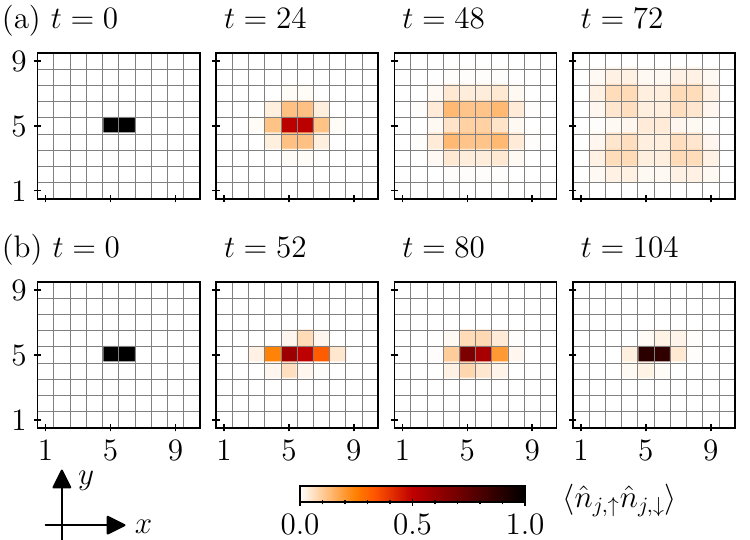}
\caption{Snapshots of doublon density for the system on a square lattice at different time for (a) $F_{x}=F_{y}=0$, and (b) $F_{x}=0.03$, $F_{y}=0.06$. 
Other parameters of the system are $L_{x}=10$, $L_{y}=9$, $U=-10$, $V=(J_{\uparrow }^{2}+J_{\downarrow }^{2})/U=-0.101$, $J_{\downarrow}=1$ and $J_{\uparrow}=0.1$.}
\label{fig_DoublonsBOs2D}
\end{figure}

\section{Numerical results of two-dimensional system}
\label{BOs2D}

So far, we have clarified the mechanism of doublons BOs in the $1$D mass-imbalanced extended Hubbard model. Since the doublons effective Hamiltonian is independent of spatial dimension, the generalization of doublons BOs to a higher dimension is straightforward for the resonance case $V=(J_{\uparrow }^{2}+J_{\downarrow }^{2})/U$. Here we perform numerical simulations of time evolutions for the $2$D system. The Hubbard model is taken as the Hamiltonian Eq. (\ref{H_hubbard}) on a square lattice with size $(L_{x},L_{y})=(10,9)$. While the $2$D tilted potential \cite{kolovsky2003bloch, witthaut2004bloch} is 
\begin{equation}
\hat{H}_{\mathrm{dc}}^{(\mathrm{2D})}=\sum_{m,n}\mathbf{F} \cdot \mathbf{r}_{m,n}\hat{n}_{(m,n)},
\end{equation}%
where $\mathbf{F}=(F_{x}, F_{y})$ is the strength of $2$D tilted potential, and $\mathbf{r}_{m,n}=m\mathbf{x}+n\mathbf{y}$ is the lattice vector of the square lattice with unit vectors $\mathbf{x}$ and $\mathbf{y}$ along the $x$-axis and $y$-axis, respectively.

Taking two NN doublons localized in the center of the lattice as an initial state, Figs. \ref{fig_DoublonsBOs2D}(a) and \ref{fig_DoublonsBOs2D}(b) show the numerical results of doublon density at different time for zero and nonzero  tilted potential $\mathbf{F}$, respectively. As expected, Fig. \ref{fig_DoublonsBOs2D}(a) shows that the doublons manifest the spreading dynamics resembling the free particles in a square lattice. While the doublons in Fig. \ref{fig_DoublonsBOs2D}(b) undergo the BOs, with periods $T_{x}=2\pi/(2F_{x})=104.72$ and $T_{y}=2\pi/(2F_{y})=52.36$ in $x$ and $y$ directions, respectively.

\section{Conclusion}

\label{conclusion}

We have studied multi-doublon BOs in the extended Fermi-Hubbard model with imbalance mass. The doublons effective Hamiltonian is given to elucidate the mechanisms of two types of doublons BOs, i.e., the bound-doublon and free-doublon BOs. The interaction between doublons stems not only from the NN interaction between fermions but also from the hopping and on-site interaction. Therefore, there is a resonance point in the system parameters that results in the free-doublon dynamics. It is shown that the spin-up and spin-down fermions are always well bounded as doublons and exhibit two types of BOs dynamics, despite the imbalance of their mass. Numerical results of the time evolution of doublon density and MSD demonstrate our conclusions. The numerical simulations of the system on a square lattice suggest that the analysis can be applied to higher dimensions. 
Although the presence of mass imbalance is not essential for the observation of doublons BOs, in contrast to the nearest-neighbor interaction, it may facilitates the realization in various experimental platforms. 
Our findings highlight the role of NN interaction in the nonequilibrium dynamic properties, as well as pave the way for the future explorations of BOs in strongly correlated systems.

\acknowledgments This work was supported by Research Fund of Guangdong-HongKong-Macao Joint Laboratory for Intelligent Micro-Nano Optoelectronic Technology (No. 2020B1212030010).

\label{A} \setcounter{equation}{0} \renewcommand{\theequation}{A%
\arabic{equation}}
\label{A} \setcounter{figure}{0} \renewcommand{\thefigure}{A%
\arabic{figure}}

\setcounter{section}{0} \renewcommand{\thesection}{APPENDIX A}
\section{Derivation of the doublons effective Hamiltonian}
\label{appendix_a}

Following the previous works on the $J/U$ perturbation expansion of the
Fermi-Hubbard model using the Schrieffer-Wolff transformation \cite{chao1977degenerate, chao1977kinetic, chao1978canonical, macdonald1988t}, we present
a detailed derivation of the doublons effective Hamiltonian Eq. (\ref{H_eff}%
) in the strong on-site interaction region $\left\vert U\right\vert \gg
J_{\sigma },\left\vert V\right\vert ,\left\vert F\right\vert $. In this
case, the states with sites being either doubly occupied or empty are well
separated from others in the energy spectrum.

\begin{figure}[h]
\centering
\includegraphics[width=0.5\textwidth]{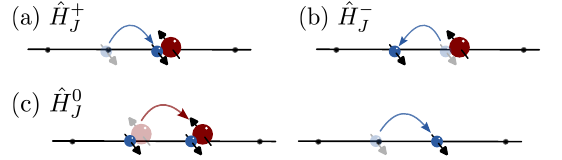}
\caption{Schematic illustrations of  three types of hopping processes described by (a) $\hat{H}_{J}^{+}$, (b) $\hat{H}_{J}^{-}$ and (c) $\hat{H}_{J}^{0}$.}
\label{fig_hopping}
\end{figure}

The Hamiltonian in Eq. (\ref{H_hubbard}) can be divided into five parts 
\begin{equation}
\hat{H}=\hat{H}_{J}^{+}+\hat{H}_{J}^{-}+\hat{H}_{J}^{0}+\hat{H}_{U}+\hat{H}%
_{V},
\end{equation}%
where $\hat{H}_{J}^{+}\ (\hat{H}_{J}^{-})$ describes the processes that
increase (decrease) the number of doubly occupied sites by one (see Fig. \ref{fig_hopping}), that are
written as 
\begin{eqnarray}
\hat{H}_{J}^{+} &=&-\sum_{\left\langle i,j\right\rangle }\sum_{\sigma
=\uparrow ,\downarrow }J_{\sigma }[\hat{n}_{i,\bar{\sigma}}\hat{c}_{i,\sigma
}^{\dagger }\hat{c}_{j,\sigma }(1-\hat{n}_{j,\bar{\sigma}})  \notag \\
&&+\hat{n}_{j,\bar{\sigma}}\hat{c}_{j,\sigma }^{\dagger }\hat{c}%
_{i,\sigma }(1-\hat{n}_{i,\bar{\sigma}})],
\end{eqnarray}%
and $\hat{H}_{J}^{-}=(\hat{H}_{J}^{+})^{\dagger }$, in which $\bar{\sigma}$
denotes spin in the opposite direction of spin $\sigma $. While $\hat{H}%
_{J}^{0}$ describes the processes that do not change the number of doubly
occupied sites, which has the form 
\begin{eqnarray}
\hat{H}_{J}^{0} &=&-\sum_{\left\langle i,j\right\rangle }\sum_{\sigma
=\uparrow ,\downarrow }J_{\sigma }[\left( 1-\hat{n}_{i,\bar{\sigma}}\right) 
\hat{c}_{i,\sigma }^{\dagger }\hat{c}_{j,\sigma }\left( 1-\hat{n}_{j,\bar{%
\sigma}}\right)  \notag \\
&&+\hat{n}_{i,\bar{\sigma}}\hat{c}_{i,\sigma }^{\dagger }\hat{c}_{j,\sigma }%
\hat{n}_{j,\bar{\sigma}}+\mathrm{H.c.}].
\end{eqnarray}%
The on-site and NN interaction terms are $\hat{H}_{U}=U\sum_{i}%
\hat{n}_{i,\uparrow }\hat{n}_{i,\downarrow }$ and $\hat{H}_{V}=
(V/2)\sum_{\left\langle i,j\right\rangle }\hat{n}_{i}\hat{n}_{j}$,
respectively.

We assume that the on-site interaction is strong, that is $\left\vert
U\right\vert \gg J_{\sigma },$ $\left\vert U\right\vert\gg \left\vert
V\right\vert,$ and perform the Schrieffer-Wolff transformation 
\begin{eqnarray}
\hat{H}_{\mathrm{eff}} &=&e^{\mathrm{i}\hat{S}}\hat{H}e^{-\mathrm{i}\hat{S}}
\notag \\
&=&\hat{H}+\mathrm{i}\left[ \hat{S},\hat{H}\right] +\frac{\mathrm{i}^{2}}{2}%
\left[ \hat{S},\left[ \hat{S},\hat{H}\right] \right] +...  \notag \\
&=&\hat{H}_{U}+\hat{H}_{V}+\hat{H}_{J}^{0}+\hat{H}_{J}^{+}+\hat{H}_{J}^{-}+%
\mathrm{i}\left[ \hat{S},\hat{H}_{U}\right]  \notag \\
&&+\mathrm{i}\left[ \hat{S},\hat{H}_{V}\right] +\mathrm{i}\left[ \hat{S},%
\hat{H}_{J}^{+}+\hat{H}_{J}^{-}\right] +\mathrm{i}\left[ \hat{S},\hat{H}%
_{J}^{0}\right]  \notag \\
&&+\frac{\mathrm{i}^{2}}{2}\left[ \hat{S},\left[ \hat{S},\hat{H}\right] %
\right] +...,
\end{eqnarray}%
where $\hat{S}$ requires to be a Hermitian
operator, so that the above transformation is unitary.  We would like to keep the term in $\hat{H}_{\mathrm{eff}}$ up to $O(J^{2}/U)$. This can be done by taking $\hat{S}$ up to second order, that is
\begin{equation}
	\hat{S}=\hat{S}^{(1)}+\hat{S}^{(2)},
\end{equation}
 where $\hat{S}^{(1)}=O(J/U)$ and $\hat{S}^{(2)}=O(J^{2}/U^{2})$. Here we set $O(J_{\uparrow })=O(J_{\downarrow })=O(J).$ One can check that $ \hat{H}_{U}$ and $ \hat{H}_{\lambda}$ satisfy
 \begin{equation}
 	\left[ \hat{H}_{U}, \hat{H}%
_{J}^{\lambda }\right] =\lambda U\hat{H}_{J}^{\lambda }, 
\label{commu_HUHJ}
 \end{equation}
where $\lambda=0,\pm$. This commutation relation is useful for the upcoming discussion.

Obviously, the largest term in $\left[ \hat{S},\hat{H}\right]$ is $\left[ \hat{S},\hat{H}_U\right]$, which can be exploited to cancel other terms. 
We firstly wish to eliminate the term $\hat{H}_{J}^{+}+\hat{H}_{J}^{-},$ which can be done by
selecting $\hat{S}^{(1)}$ such that 
\begin{equation}
	\hat{H}_{J}^{+}+\hat{H}_{J}^{-}+%
\mathrm{i}\left[ \hat{S}^{(1)},\hat{H}_{U}\right] =0.
\label{HJSHU}
\end{equation}
Using Eq. (\ref{commu_HUHJ}), we can check that 
\begin{equation}
\hat{S}^{(1)}=-\frac{\mathrm{i}}{U}\left( \hat{H}_{J}^{+}-\hat{H}_{J}^{-}\right) 
\end{equation}%
does satisfy Eq. (\ref{HJSHU}).

To eliminate the term $\left[ \hat{S}^{(1)},\hat{H}_{J}^{0}\right]$, we consider the second order term of $\hat{S}$, which is determined by the
operator equation 
\begin{equation}
	\mathrm{i}\left[ 
\hat{S}^{(1)},\hat{H}_{J}^{0}\right]+\mathrm{i}\left[ \hat{S}^{(2)},\hat{H}_{U}\right] =0.
\label{op_eq}
\end{equation}
Again, using Eq. (\ref{commu_HUHJ}) and the Jacobi identity, we can check that 
\begin{equation}
	\hat{S}^{(2)}=-\frac{\mathrm{i}}{U^2}\left[\hat{H}_{J}^{+}+\hat{H}_{J}^{-},\hat{H}_{J}^{0} \right]
\end{equation}
satisfy Eq. (\ref{op_eq}). Thus the terms $\left[ \hat{S}^{(2)},\hat{H}_{J}^{0}\right]$ and $\left[ \hat{S}^{(2)},\hat{H}_{J}^{+}+\hat{H}_{J}^{-}\right]$ are of the order of $J^{3}/U^{2}$, and term $\left[ \hat{S}^{(2)},\hat{H}_{V}\right]$ is of the order of  $J^{2}V/U^{2}$.

In addition, direct calculations yield
\begin{equation}
	\textrm{i}\left[ \hat{S}^{(1)},\hat{H}_{J}^{+}+\hat{H}_{J}^{-}\right]=\frac{2}{U}\left[ \hat{H}_{J}^{+},\hat{H}_{J}^{-}\right],
\end{equation}
and
\begin{equation}
	\frac{\mathrm{i}^{2}}{2}\left[ \hat{S}^{(1)},\left[ \hat{S}^{(1)},\hat{H}_{U}\right]\right]=-\frac{1}{U}\left[ \hat{H}_{J}^{+},\hat{H}_{J}^{-}\right].
\end{equation}
Other terms in $\mathrm{i}^{2}\left[ \hat{S},\left[ \hat{S},\hat{H}\right]\right]/2$ have the order that higher than $J^{2}/U$.

Therefore, we obtain 
\begin{eqnarray}
\hat{H}_{\mathrm{eff}} &=&\hat{H}_{U}+\hat{H}_{V}+\hat{H}_{J}^{0}+\mathrm{i}%
\left[ \hat{S}^{(1)},\hat{H}_{V}\right] +\frac{1}{U}\left[ \hat{H}_{J}^{+},\hat{H}%
_{J}^{-}\right]  \notag \\
&&+O\left(J^{2}V/U^{2}\right) +O\left( J^{3}/U^{2}\right) .
\label{H_eff_complex}
\end{eqnarray}

In fact, the energy scale $J^2/U$ in Eq. (\ref{H_eff_complex}) can be grasped from the second order process of degenerate perturbation theory, resembling Eq. (\ref{perturbation_H}). For example, when taking the doublons states as the degenerate basis, and the hopping term as perturbation, the hopping of a doublon is achieved by the hopping of two fermions through the second order process, which includes other states possessing energy gaps from the doublons states of order $U$, as an intermediate process. Thus, this process possesses  energy scale $J^2/U$. 
Due to a large $\left\vert U\right\vert $, the doublons states are well
separated from others with the energy gaps of order $\left\vert U\right\vert 
$. Then the Hamiltonian can be safely projected into the doublons
subspace up to order $J^{2}/U$. The effective Hamiltonian in Eq. (\ref{H_eff_complex}) can be
simplified by taking the following consideration into account: (i) Term $[\hat{%
S}^{(1)},\hat{H}_{V}]$ can be omitted, since it connects the doublons subspace
to others. (ii) Term $\hat{H}_{J}^{0}$ describing the hopping
processes between singly occupied sites and empty (or doubly occupied)
sites, also vanishes in the doublons subspace.

We finally need to evaluate the commutator $\left[ \hat{H}_{J}^{+},\hat{H}%
_{J}^{-}\right] .$ Since $\hat{H}_{J}^{+}$ and $\hat{H}_{J}^{-}$ are sums of
pair operators, nonvanishing contributions arise only if the two pairs have
one or two sites in common. Denoting $\hat{H}_{J}^{\pm }=\sum_{\left\langle
i,j\right\rangle }\hat{H}_{J,(i,j)}^{\pm }$, we have 
\begin{eqnarray}
&&\left[ \hat{H}_{J}^{+},\hat{H}_{J}^{-}\right] =\sum_{\left\langle
i,j\right\rangle }\sum_{\left\langle m,n\right\rangle }\left[ \hat{H}%
_{J,(i,j)}^{+},\hat{H}_{J,(m,n)}^{-}\right]   \\
&=&\sum_{\left\langle i,j\right\rangle }\left[ \hat{H}_{J,(i,j)}^{+},\hat{H}%
_{J,(i,j)}^{-}\right] +\sum_{\left\langle i,j,m\right\rangle }\left[ \hat{H}%
_{J,(i,j)}^{+},\hat{H}_{J,(j,m)}^{-}\right] , \notag
\end{eqnarray}%
where $\sum_{\left\langle i,j,m\right\rangle }$ denotes the summation over
trimers with site $j$ being a nearest neighbor of both sites $i$ and $m$.
It can be checked that these three-site processes vanish in the doublons
subspace, and the two-site processes remain. Direct calculation shows that 
\begin{eqnarray}
&&\sum_{\left\langle i,j\right\rangle }\left[ \hat{H}_{J,(i,j)}^{+},\hat{H}%
_{J,(i,j)}^{-}\right]   \notag \\
&=&(J_{\uparrow }^{2}+J_{\downarrow }^{2})\left[ 2\sum_{i}\hat{n}%
_{i,\uparrow }\hat{n}_{i,\downarrow }-\sum_{\left\langle i,j\right\rangle }(%
\hat{n}_{i,\uparrow }\hat{n}_{j,\downarrow }+\hat{n}_{i,\downarrow }\hat{n}%
_{j,\uparrow })\right]   \notag \\
&&+2J_{\uparrow }J_{\downarrow }\sum_{\left\langle i,j\right\rangle }(\hat{d}%
_{i}^{\dagger }\hat{d}_{j}+\hat{d}_{j}^{\dagger }\hat{d}_{i}+\hat{s}_{i}^{+}%
\hat{s}_{j}^{-}+\hat{s}_{j}^{+}\hat{s}_{i}^{-}),
\label{commu}
\end{eqnarray}%
where $\hat{s}_{i}^{+}=\hat{c}_{i,\uparrow }^{\dagger }\hat{c}_{i,\downarrow
}$, $\hat{s}_{i}^{-}=(\hat{s}_{i}^{+})^{\dagger }$ are the spin-flip
operator, and $d_{i}^{\dagger }=\hat{c}_{i,\uparrow }^{\dagger }\hat{c}%
_{i,\downarrow }^{\dagger }$ is the creation operator of doublon. Note that
in the doublons subspace, terms $\hat{s}_{i}^{+}\hat{s}_{j}^{-}$ and $\hat{s}%
_{i}^{z}\hat{s}_{j}^{z}=(\hat{n}_{i,\uparrow }-\hat{n}_{i,\downarrow })(\hat{%
n}_{j,\uparrow }-\hat{n}_{j,\downarrow })/4$ vanish, and the on-site
interacting term gives constant energy for a fixed number of doublons. Then term $-2\hat{s}_{i}^{z}\hat{s}%
_{j}^{z}$ can be added to Eq. (\ref{commu}). Up to order $J^{2}/U$, the effective
Hamiltonian that conserves the total number of doublons is 
\begin{equation}
\hat{H}_{\mathrm{eff}}=\sum_{\left\langle i,j\right\rangle }\left[ J_{%
\mathrm{eff}}\left( \hat{d}_{i}^{\dagger }\hat{d}_{j}+\hat{d}_{j}^{\dagger }%
\hat{d}_{i}\right) +\frac{1}{2}V_{\mathrm{eff}}\hat{n}_{i}\hat{n}_{j}\right]
,
\end{equation}%
where the effective hopping amplitude and NN interaction of
doublons are $J_{\mathrm{eff}}=2J_{\uparrow }J_{\downarrow }/U$ and $V_{%
\mathrm{eff}}=V-(J_{\uparrow }^{2}+J_{\downarrow }^{2})/U$, respectively,
and the constant term is omitted.

\setcounter{section}{0} \renewcommand{\thesection}{APPENDIX B}
\section{Chebyshev expansion method for time evolution}
\label{appendix_b}

For clarity, here we formulate the implementation of the Chebyshev expansion method for numerical
simulations of time evolution $\left|\Psi\left(t\right)\right\rangle=\exp(-\mathrm{i}\hat{\mathcal{H}}t)\left|\Psi\left(0\right)\right\rangle $.  The more detailed descriptions for this method can be found in Refs. \cite{tal1984accurate, fehske2009numerical, weisse2006kernel}.

To expand the time evolution operator into a finite series of Chebyshev polynomials, we first need to rescale the Hamiltonian
\begin{equation}
	\widetilde{\mathcal{H}}=\frac{\hat{\mathcal{H}}-b}a ,
\end{equation}
fitting the spectrum into the interval $\left[-1, 1 \right]$. The scaling factors are $a=(E_\text{max}{-}E_\text{min})/(2-\epsilon)$ and $b=(E_\text{max}{+}E_\text{min})/2$, where we take a small cutoff $\epsilon=0.01$ to ensure the rescaled spectrum lies inside $\left[-1, 1 \right]$, and the extremal eigenvalues of the Hamiltonian, $E_\text{max}$ and $E_\text{min}$, can be efficiently computed by the the implicitly restarted Arnoldi method \cite{lehoucq1998arpack}. Then the time evolution operator can be expanded as a series of Chebyshev polynomials
\begin{eqnarray}
	&&\exp(-\mathrm{i}\hat{\mathcal{H}}t)\left|\Psi\left(0\right)\right\rangle \label{Cheb_exp}\\
	&&\approx \exp(-\mathrm{i}bt)\left[J_{0}(at)+2\sum_{n=1}^{N}(-\mathrm{i})^{n}J_{n}(at)T_n(\widetilde{\mathcal{H}})\right]\left|\Psi\left(0\right)\right\rangle , \notag
\end{eqnarray}
where $J_{n}(x)$ denotes the first-kind Bessel function of order $n$, and $T_n(x)=\cos(n \arccos(x))$ denotes the first-kind Chebyshev polynomials of order $n$. The term $ |\nu_{n}\rangle =T_n(\widetilde{\mathcal{H}})\left|\Psi\left(0\right)\right\rangle$ in Eq. (\ref{Cheb_exp}) can be computed iteratively by the following recurrence relation of the Chebyshev polynomials
\begin{eqnarray}
	&& |\nu_{n+1}\rangle = 2\widetilde{\mathcal{H}}|\nu_{n}\rangle-|\nu_{n-1}\rangle,\notag\\
	&& |\nu_{1}\rangle =\widetilde{\mathcal{H}}|\nu_{0}\rangle,
\end{eqnarray}
where $|\nu_{0}\rangle =\left|\Psi\left(0\right)\right\rangle$. The algorithm requires memory only for a few vectors and for the sparse matrix Hamiltonian. For a large $n$, the Bessel function has the asymptotic behavior 
\begin{equation}
J_{n}(at)\sim\frac{1}{\sqrt{2\pi n}}\left(\frac{\text{e}at}{2n}\right)^{n},
\end{equation}
 which decay superexponentially for $n \gg at$, and the series in Eq. (\ref{Cheb_exp}) can be truncated with negligible error, i.e.,  
we can choose $N$ such that for $n>N$, the expansion coefficients $J_{n}(at)$ is smaller than a desired accuracy cutoff. 

The Chebyshev expansion method relies only on matrix vector multiplications, and thus can exploit the sparsity of the matrix representation of the Hamiltonian to save the computer memory and computation time significantly. In addiction, unlike the differencing method, this method does not accumulate errors, and taking a large time step is possible by increasing $N$.  It is especially suitable for the computations of time evolution of quantum state under time-independent Hamiltonian.


%\bibliography{DoublonsBO.bib}

\end{document}